\documentclass[11pt,twoside]{article}
\usepackage{asp2004}
\usepackage{psfig}
\usepackage{epsf}
\usepackage{graphics}
\usepackage{lscape}
\def\edcomment#1{\iffalse\marginpar{\raggedright\sl#1\/}\else\relax\fi}
\reversemarginpar

\begin{document}
\title{Automated Coronal Hole Detection using He~{\footnotesize I} 1083 nm Spectroheliograms 
and Photospheric Magnetograms}

\author{C. J. Henney and J. W. Harvey}
\affil{National Solar Observatory, Tucson, Arizona, USA}

\begin{abstract}
A method for automated coronal hole detection using He~{\footnotesize I}
1083~nm spectroheliograms and photospheric magnetograms is presented here. 
The unique line formation of the helium line allows for 
the detection of regions associated with 
solar coronal holes with minimal line-of-sight obscuration across the 
observed solar disk. The automated detection algorithm utilizes 
morphological image analysis, thresholding and smoothing to estimate
the location, boundaries, polarity and flux of candidate coronal hole
regions. The algorithm utilizes thresholds based on mean values determined
from over 10 years of the Kitt Peak Vacuum Telescope daily hand-drawn 
coronal hole images. A comparison between the automatically created and 
hand-drawn images for a 11-year period beginning in 1992 is outlined. In addition, 
the creation of synoptic maps using the daily automated coronal hole
images is also discussed.
\end{abstract}

\section{Introduction}
The association of high-speed solar wind with solar coronal holes
is of great importance for forecasting space weather events 
\citep[e.g.][]{zirk77,Arge04}. Coronal holes are low density regions 
in the solar corona that have less extreme ultraviolet and X-ray emission 
than nominal quiet and active
regions. The magnetic fields within these regions are mostly unipolar and 
extend beyond the corona into the interplanetary medium 
\citep[e.g.][]{bohlin77}. For ground-based observations, a useful proxy for 
estimating the location of coronal holes is the He~{\footnotesize I} 1083~nm line.
The unique He~{\footnotesize I} 1083~nm line formation in the chromosphere
results in decreased absorption \citep[e.g.][]{And97}.
The Kitt Peak Vacuum Telescope (KPVT) He~{\footnotesize I} 1083 nm equivalent width 
images \citep[e.g.][]{JHarvey77} are defined such that these regions appear 
bright (see Figure~1). KPVT measurements of the sun at He~{\footnotesize I} 1083~nm span
from 1974 through September 2003 \citep[e.g.][]{JHarvey1994}. A sample
KPVT He~{\footnotesize I} 1083~nm spectroheliogram, along with 
a comparison Extreme Ultraviolet Imaging 
Telescope (EIT) 19.5 nm Fe XII emission line image are shown in Figure~1
(note that the coronal hole regions appear dark in the EIT image).

\begin{figure}[!t]
\plottwo{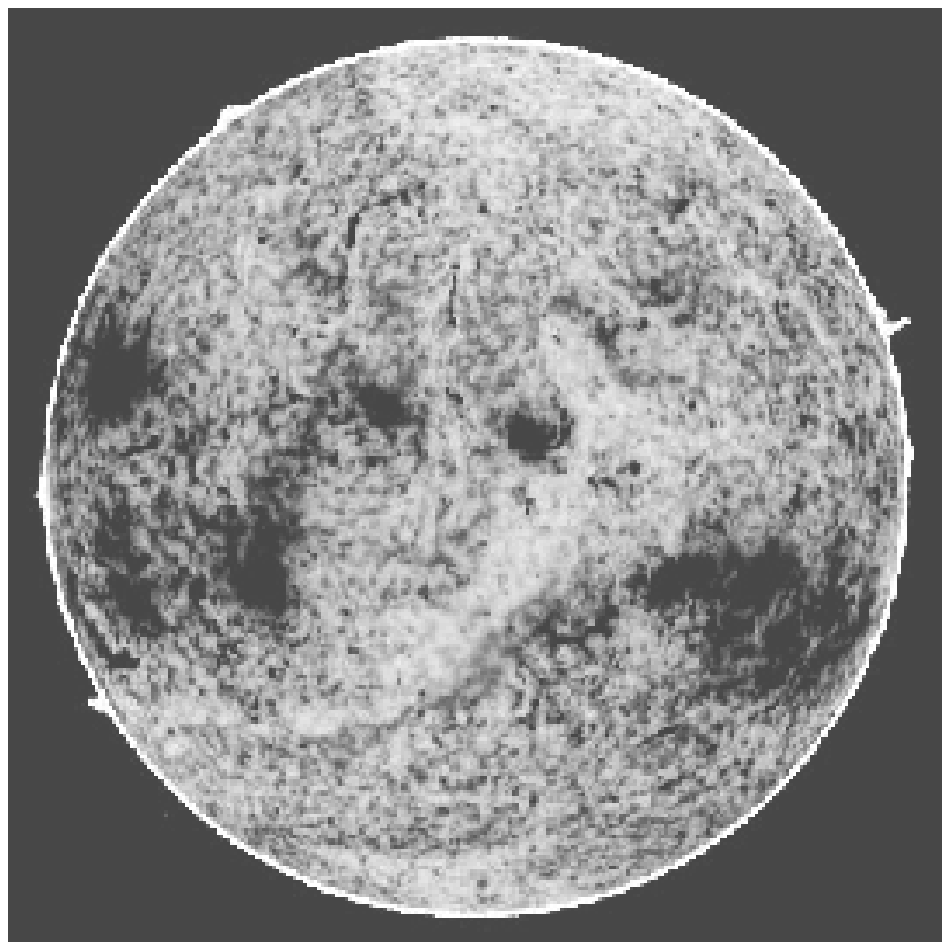}{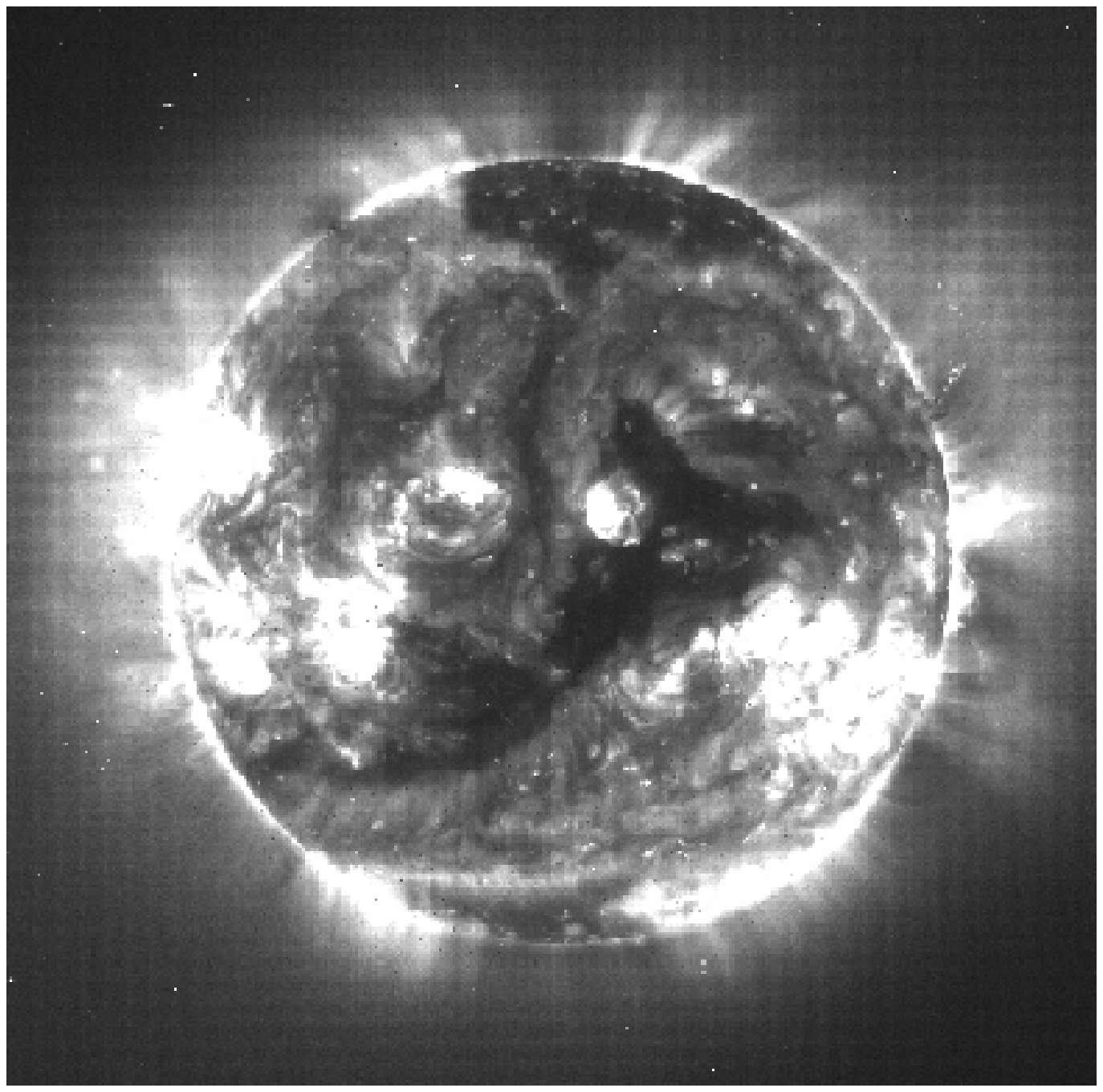}
\plottwo{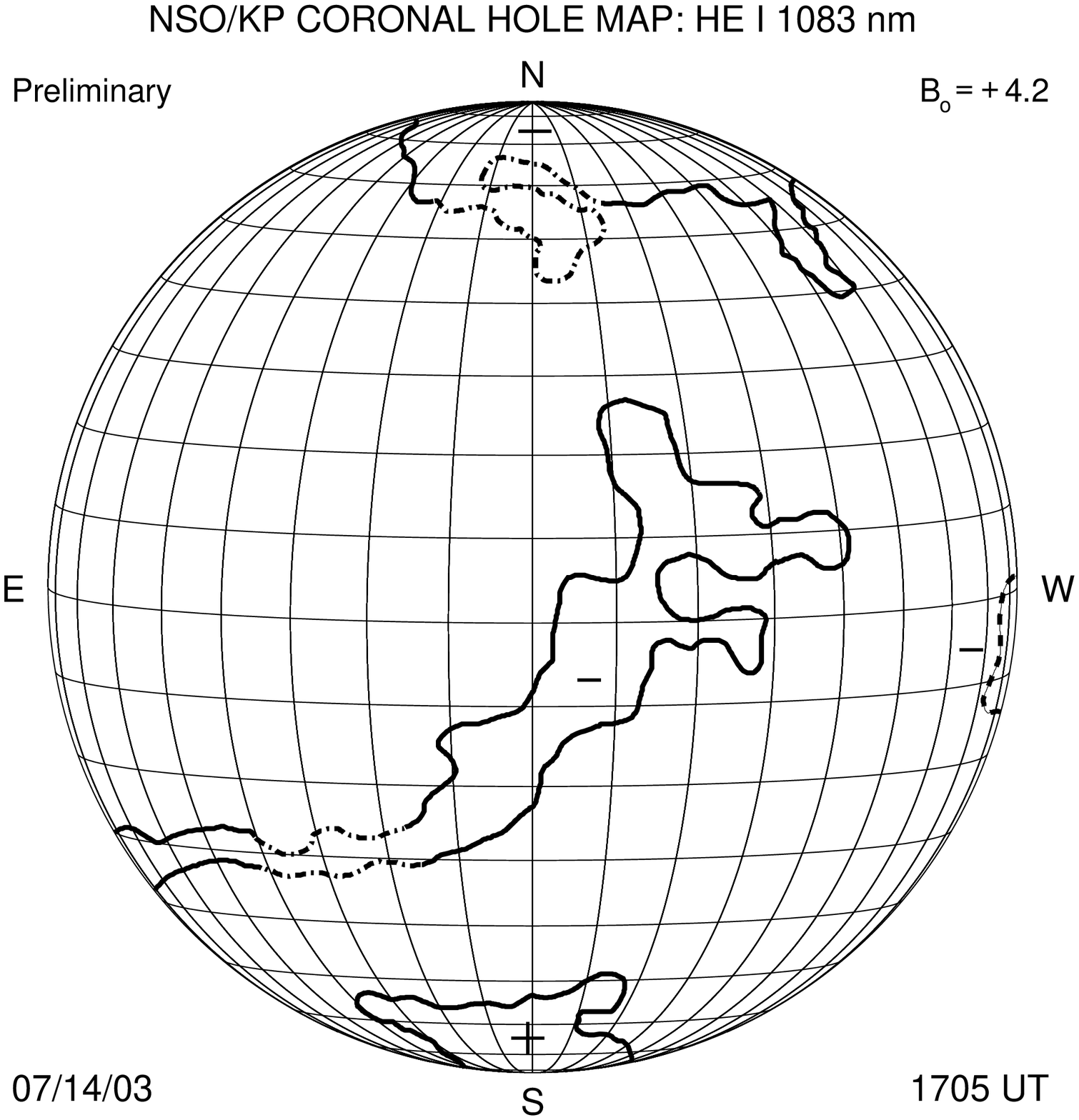}{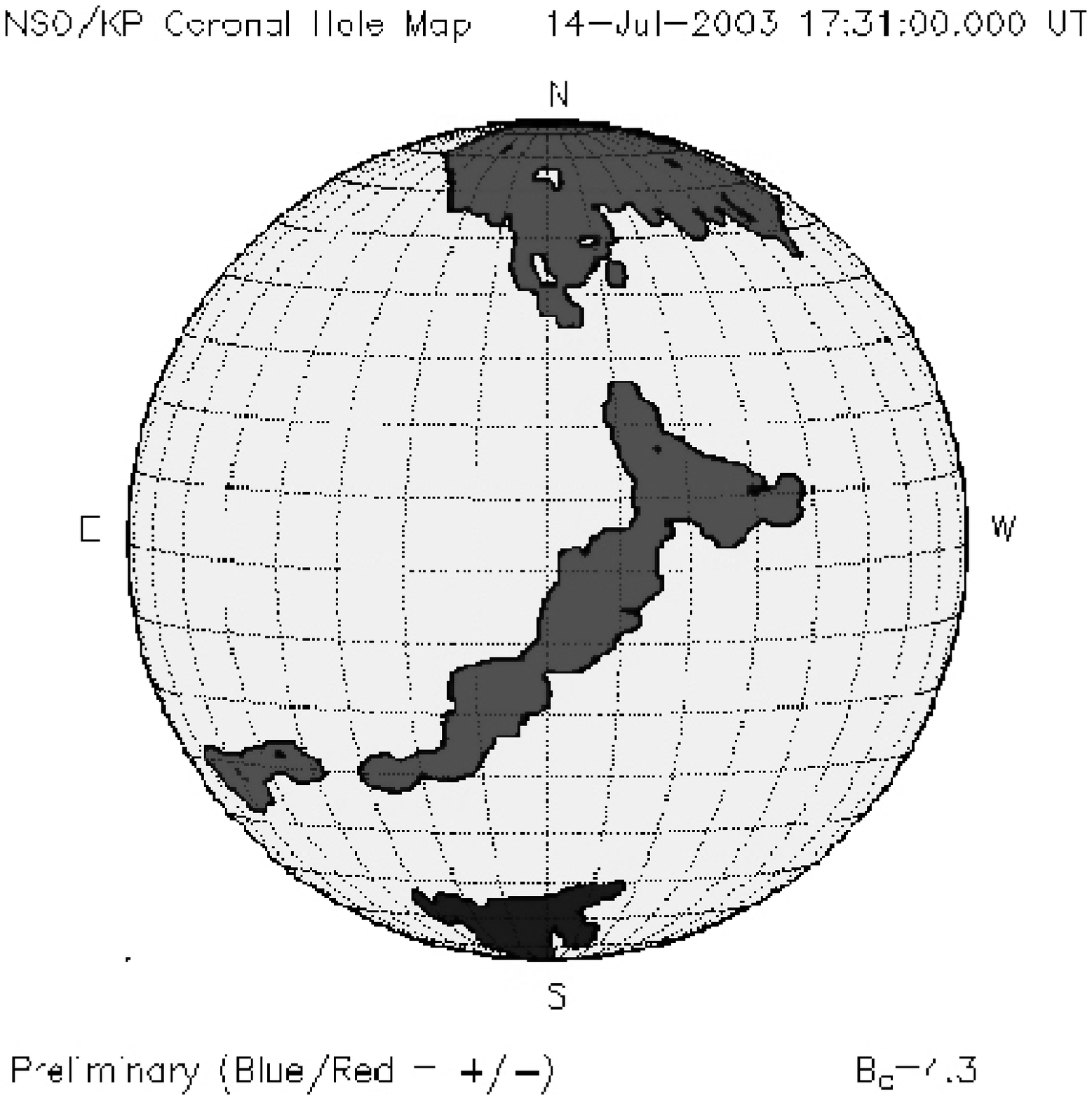}
\caption{KPVT He~{\footnotesize I} 1083~nm
spectroheliogram (top, left), EIT 19.5 nm Fe XII emission line 
image (top, right), the KPVT associated ``hand-drawn'' coronal 
hole image (bottom, left), and the automated detection coronal hole 
image (bottom, right) for July 14, 2003 at approximately 17 UT.}
\end{figure}

Using KPVT He~{\footnotesize I} 1083~nm spectroheliograms and 
photospheric magnetograms, daily computer-assisted ``hand drawn'' (HD) 
coronal hole images by Karen Harvey and Frank Recely span from April 1992 
through September 2003. In addition, they constructed coronal hole Carrington synoptic
maps for the period between September 1987 and March 2002 \citep[see][]{KHarvey2002}. 
Prior to these periods HD maps were drawn on photographic prints and incorporated
in synoptic maps published in Solar Geophysical Data. The creation
of the HD coronal hole images and synoptic maps is notably labor 
intensive, requiring iterative visual inspection of high resolution 
helium spectroheliograms and photospheric magnetograms. The production 
of HD coronal hole maps stopped with the 
conclusion of operation of the KPVT in September 2003. Unpublished 
HD daily and synoptic maps exist for periods back to 1974. 

The motivation for developing an automated detection (AD) coronal hole algorithm
is to create daily images and synoptic maps similar to the hand-drawn maps for the 
period between 1974 and 1992 using available KPVT data. In addition, the 
AD algorithm will be used to create coronal hole images using the 
daily SOLIS Vector SpectroMagnetograph (VSM) helium spectroheliograms and 
photospheric magnetograms. The SOLIS-VSM began synoptic observations in August 
2003. Initially during the development of the AD coronal hole algorithm, only the 
helium equivalent width images were utilized, at various spatial scales and 
thresholds, with limited success \citep{Henn2001}. The critical addition  
for the detection recipe presented here is the inclusion of magnetograms in the 
analysis to determine the percent unipolarity of the candidate 
region \citep[e.g.][]{KHarvey2002} to avoid areas of polarity inversions and 
centers of supergranular cells that otherwise resemble coronal holes.

Other methods for detecting coronal holes include utilizing spectral line 
properties of He~{\footnotesize I} 1083~nm \citep{Malan05} and 
multi-wavelength analysis \citep{detom05}. In future 
work, we plan to compare these methods in detail with the AD coronal hole 
algorithm presented here. The following sections outline the parameterizing 
of coronal holes (Section 2), the coronal hole detection recipe, 
along with a comparison with the HD coronal hole images and the creation 
of synoptic maps (Section 3).

\section{Coronal Hole Parameters}
The preliminary step in the development of the automated coronal
hole detection algorithm outlined in the following section is 
to parameterize unique properties of coronal holes. This allows 
for general threshold limits to be set for the input data. 
The thresholds used to select coronal hole 
candidate regions are estimated using successive pairs of KPVT 
magnetogram and helium spectroheliogram data originally used to 
create the HD coronal hole maps. Using the coronal hole boundaries
from the hand drawn images, the area of each coronal hole is projected 
into the corresponding average magnetogram and helium spectroheliogram 
image in heliographic coordinates. 

\begin{figure}[!t]
\plotone{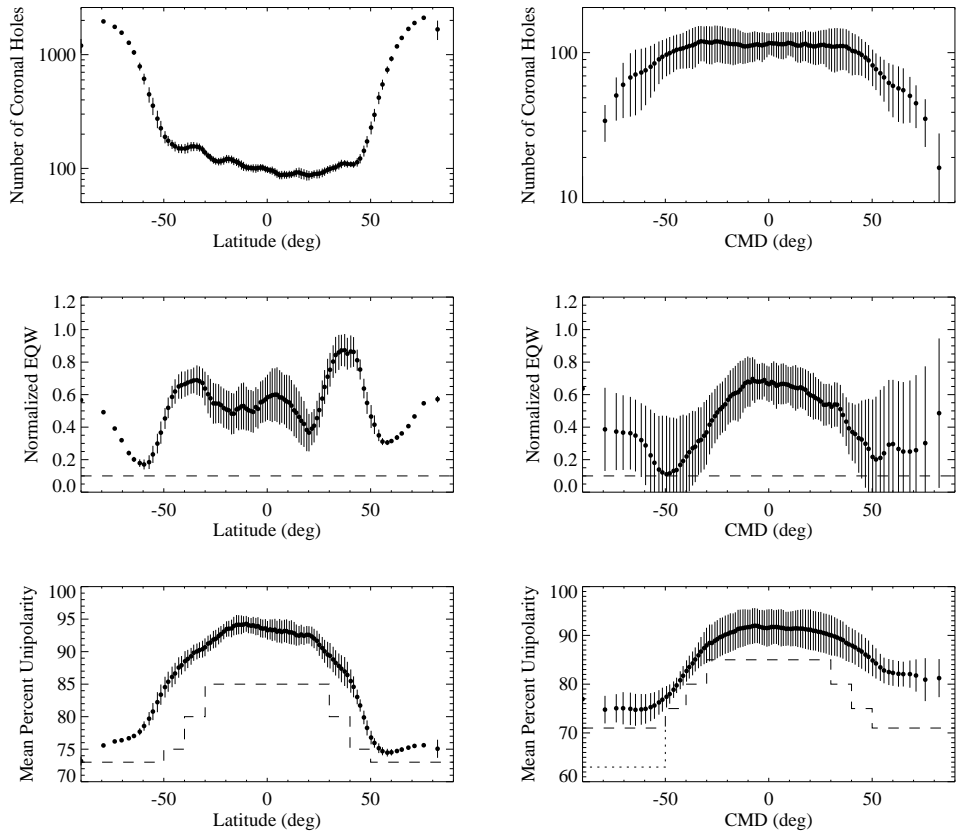}
\caption{Coronal hole parameters, for the period between 1992 through 
2003, verses latitude (left) and central meridian distance (CMD, right). 
The total number of holes (top), normalized helium 1083~nm
equivalent width (EQW, middle), and mean percent unipolarity (bottom) 
are shown. The 1-$\sigma$ error for the values is delineated by 
the vertical lines. See text for discussion.}
\end{figure}

The average images are created with 
two consecutive observations weighted inversely by one plus 
the time difference between the observations in fractional days. 
The two images are averaged in heliographic coordinates 
sine-latitude and longitude, where the older image is rotated into the time 
reference frame of the most recent image.
The KPVT data coverage for the period April 1992 through September 2003 
includes 2,781 image pairs. For this period, the boundaries of 11,241 
HD coronal holes were used to estimate the mean percent 
unipolarity for these regions relative
to latitude and central meridian distance. 

In Figure~2, the number of coronal holes (top), the normalized
He~{\footnotesize I} 1083~nm equivalent width (middle) and the mean percent 
unipolarity (bottom) are exhibited for the period between 1992 and 2003. 
The helium equivalent width is normalized by the median of all
positive values, where the zero point is essentially the mean of the disk values 
excluding active regions and filaments. Note that for the KPVT helium equivalent 
width images the limb darkening is removed. 
For this period a greater number of coronal holes were detected in
the southern hemisphere. The difference between the hemispheres could be a 
result of a delayed or slower migration of the holes to the southern polar 
region or possibly longer lived regions, since no attempt was made to 
avoid recounting long-lived regions. The greater number of detections on the 
eastern limb is most likely a result of false detections. For the 
normalized helium equivalent width and the mean unipolarity
percentage shown in Figure 2, the dashed lines delineate the threshold 
used with the AD algorithm. In addition, provisional regions are allowed for the 
east limb where the magnetic information becomes sparse due to the time difference 
between observations that constitute the average magnetogram. The threshold 
for the percent unipolarity is lowered for regions within the central meridian 
distance band between -90 and -50 degrees (depicted by the dotted line in Figure~2).

\begin{figure}
\plottwo{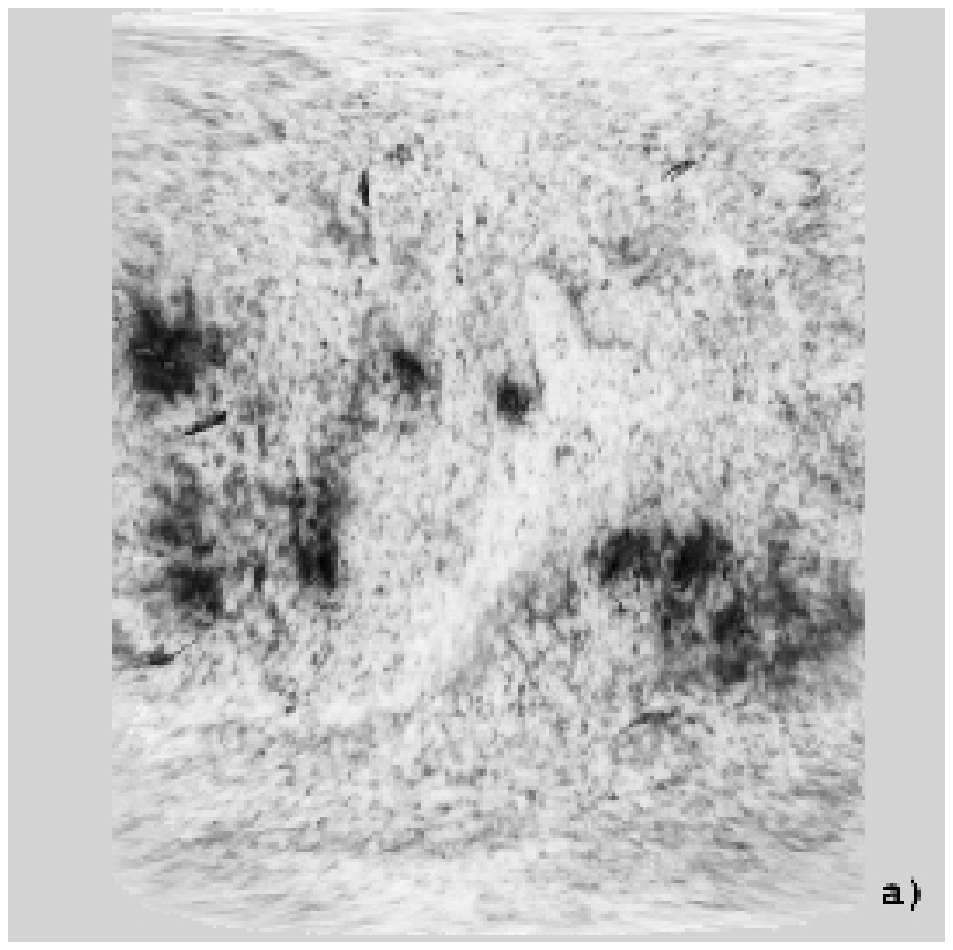}{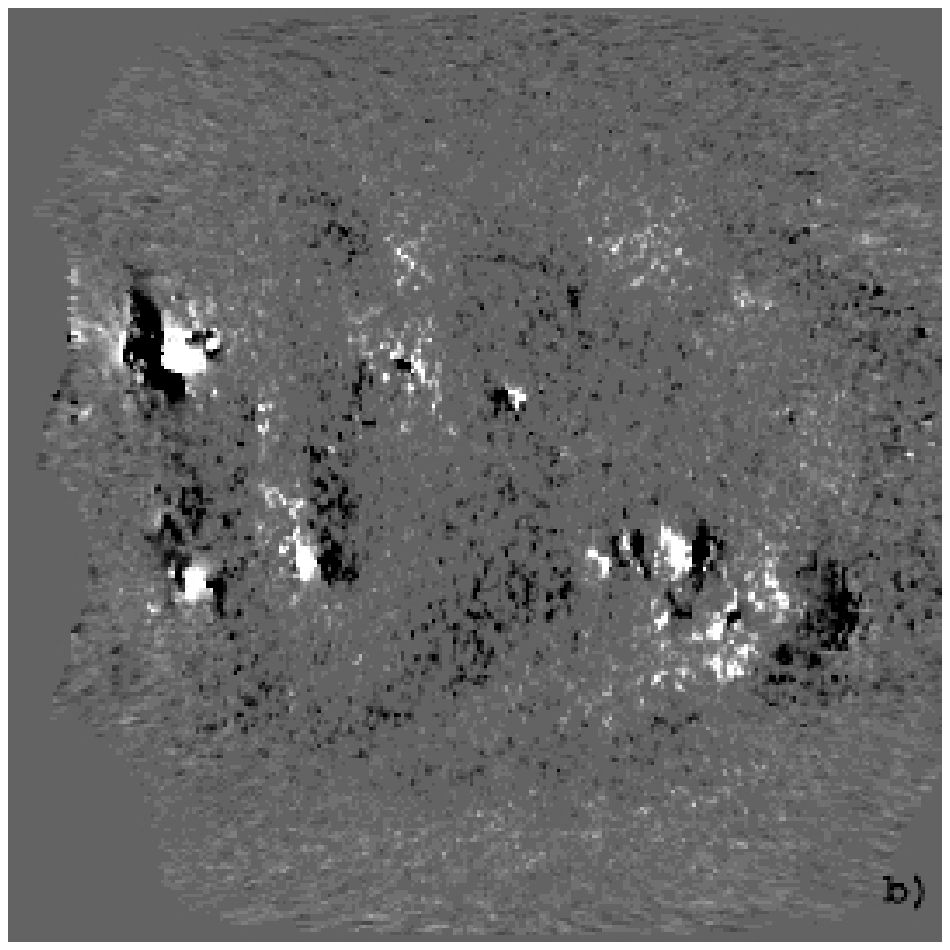}
\plottwo{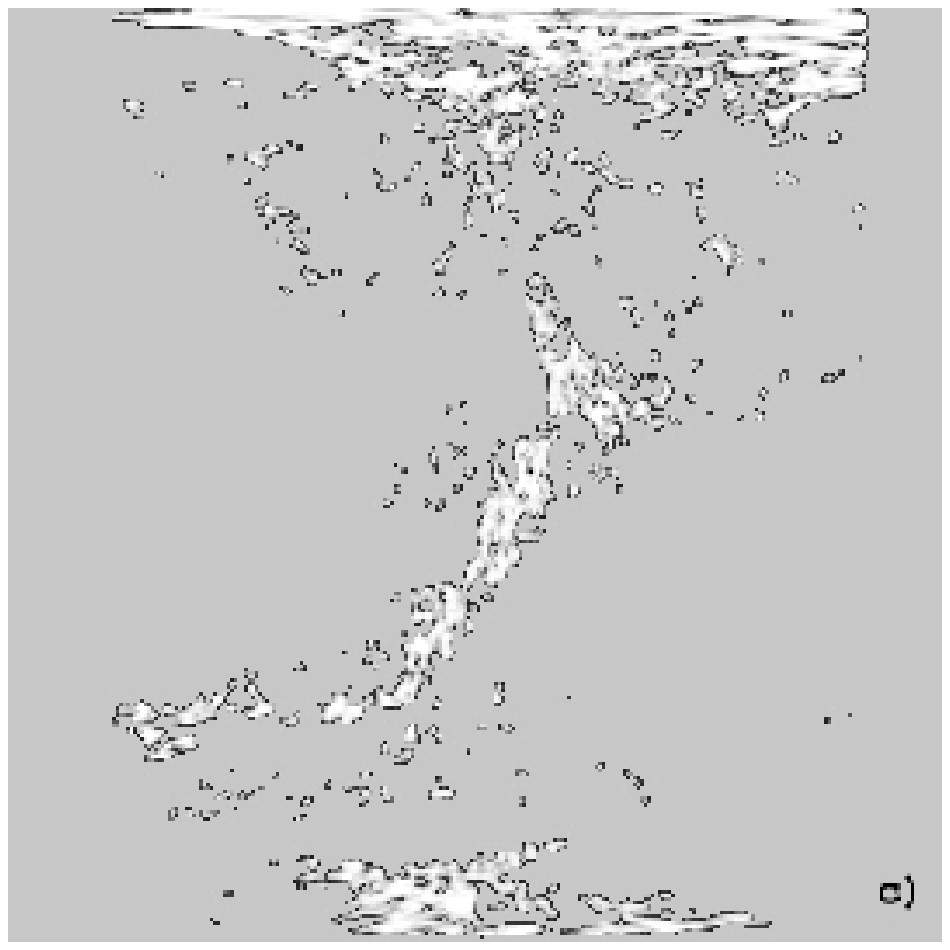}{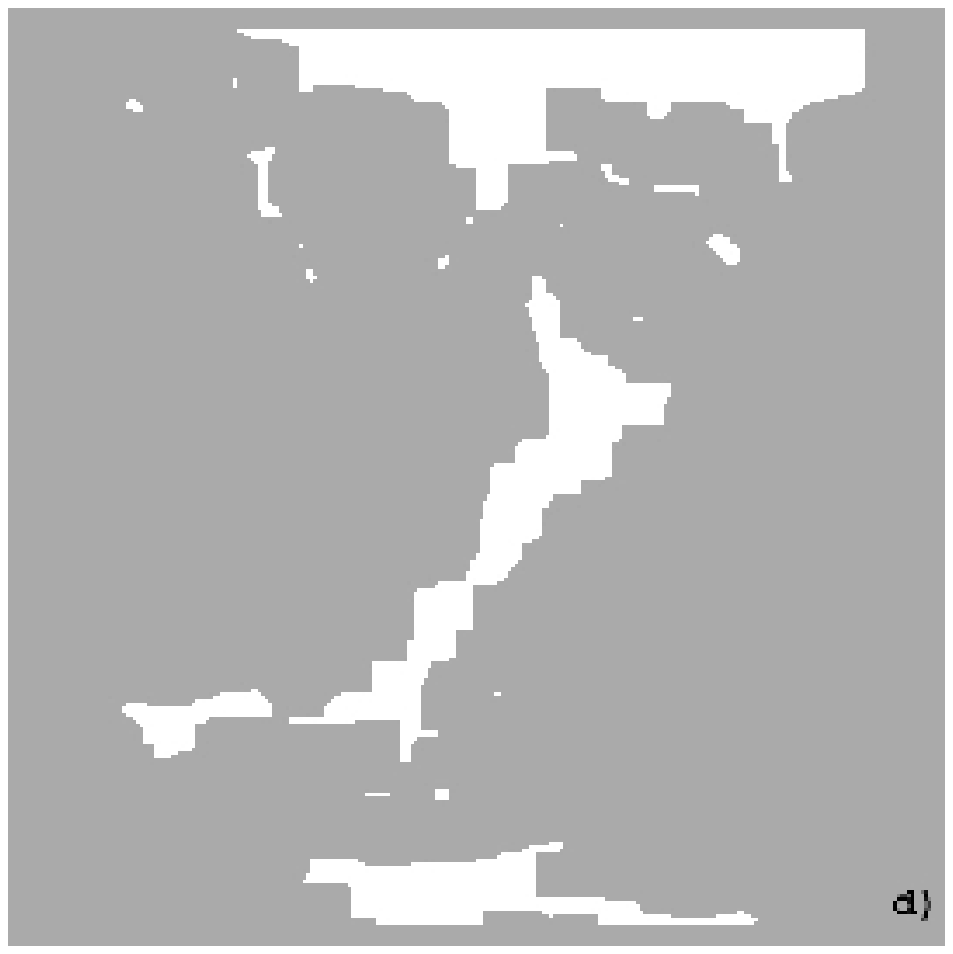}
\plottwo{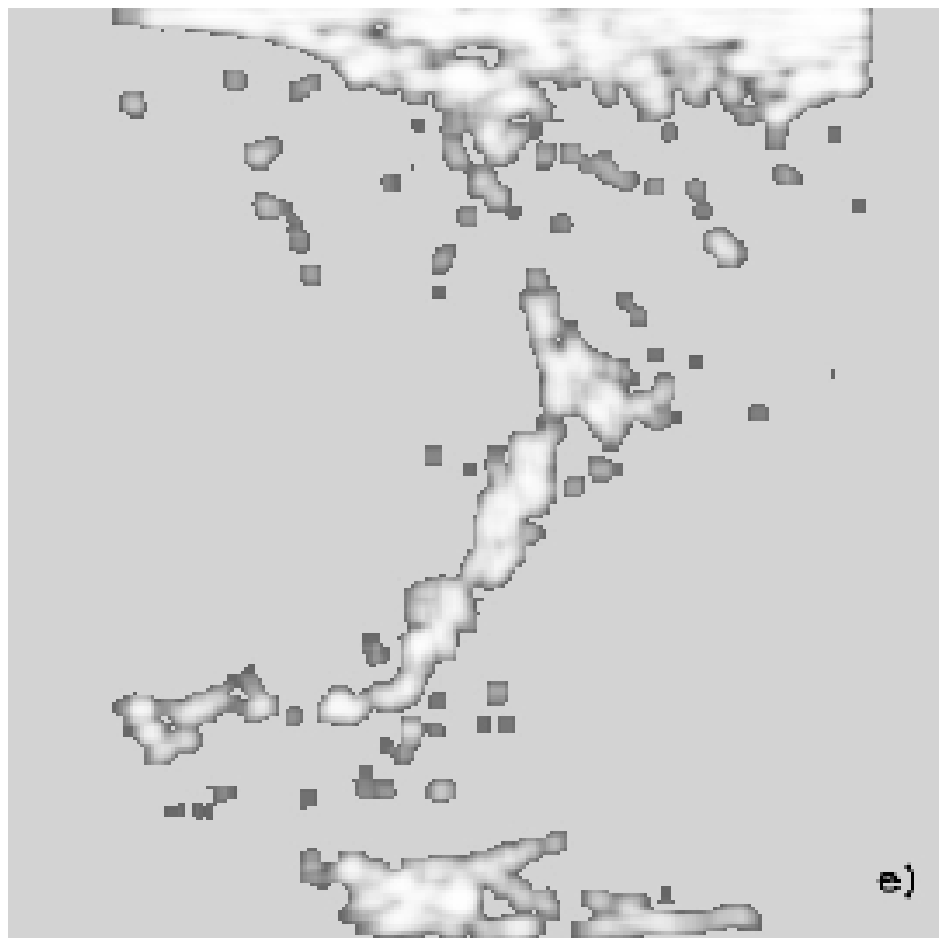}{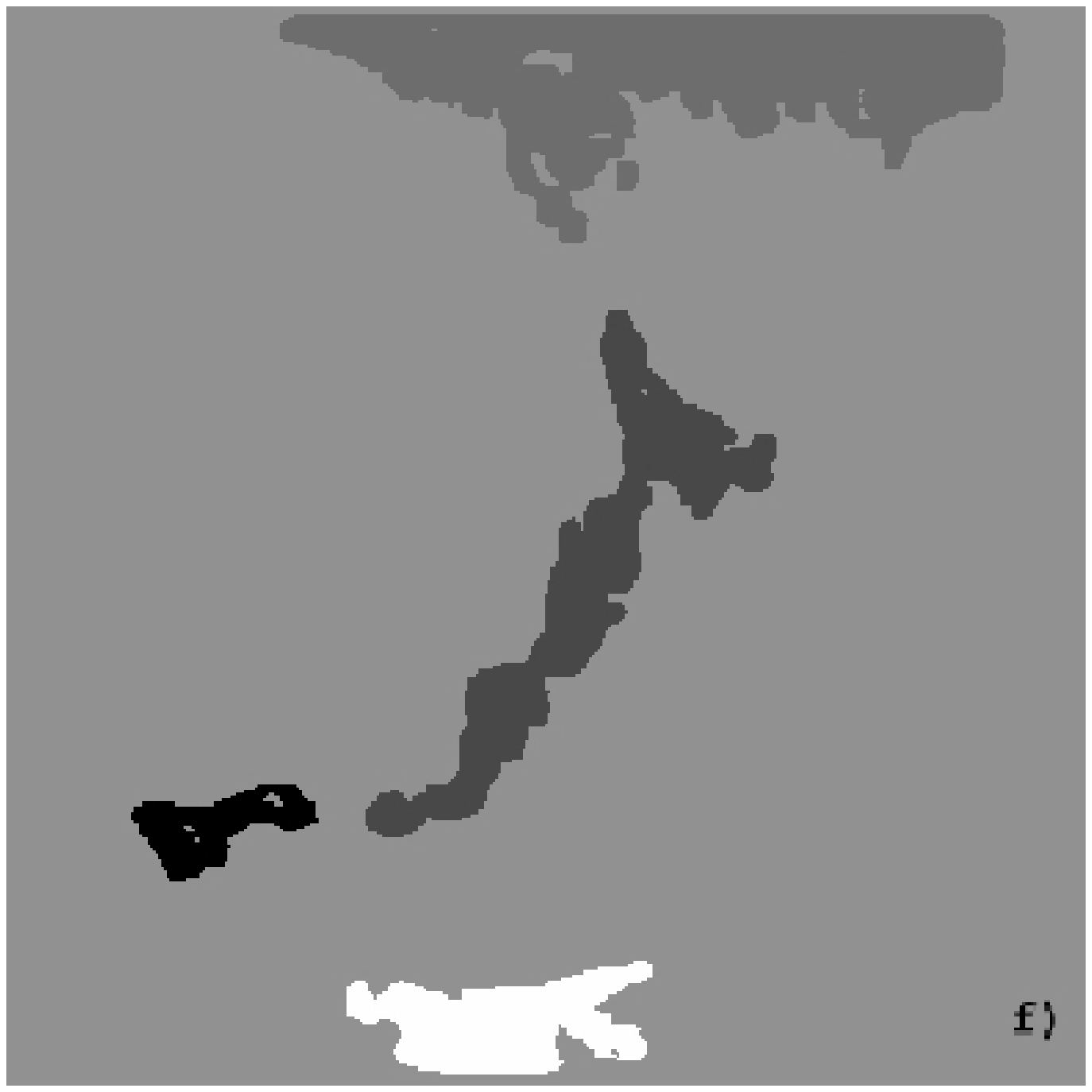}
\caption{Selected images throughout the automated coronal hole 
detection process are depicted: a) the initial average helium 
spectroheliogram in heliographic coordinates; b) the average 
magnetogram; c) after retaining only values from the image shown in a) 
above 0.10 of the median of the positive values; d) after applying the
Close function; e) after applying the Open function;
f) after removing small and low percentage unipolarity regions.}
\end{figure}

\section{Automated Detection Recipe}
The coronal hole detection process begins with a two-day average
He~{\footnotesize I} 1083~nm spectroheliogram and a two-day average 
photospheric magnetogram. The average images are created as discussed in the 
previous section. The helium image averaging assists in the detection of 
coronal holes by taking advantage of the low intrinsic network contrast 
within coronal holes \citep[e.g.][]{KHarvey2002}. The weighted average images are
trimmed on the east and west limb regions for unsigned central meridian 
distances greater than 70~degrees (see Figure~3a and~3b).

\begin{figure}[!t]
\plottwo{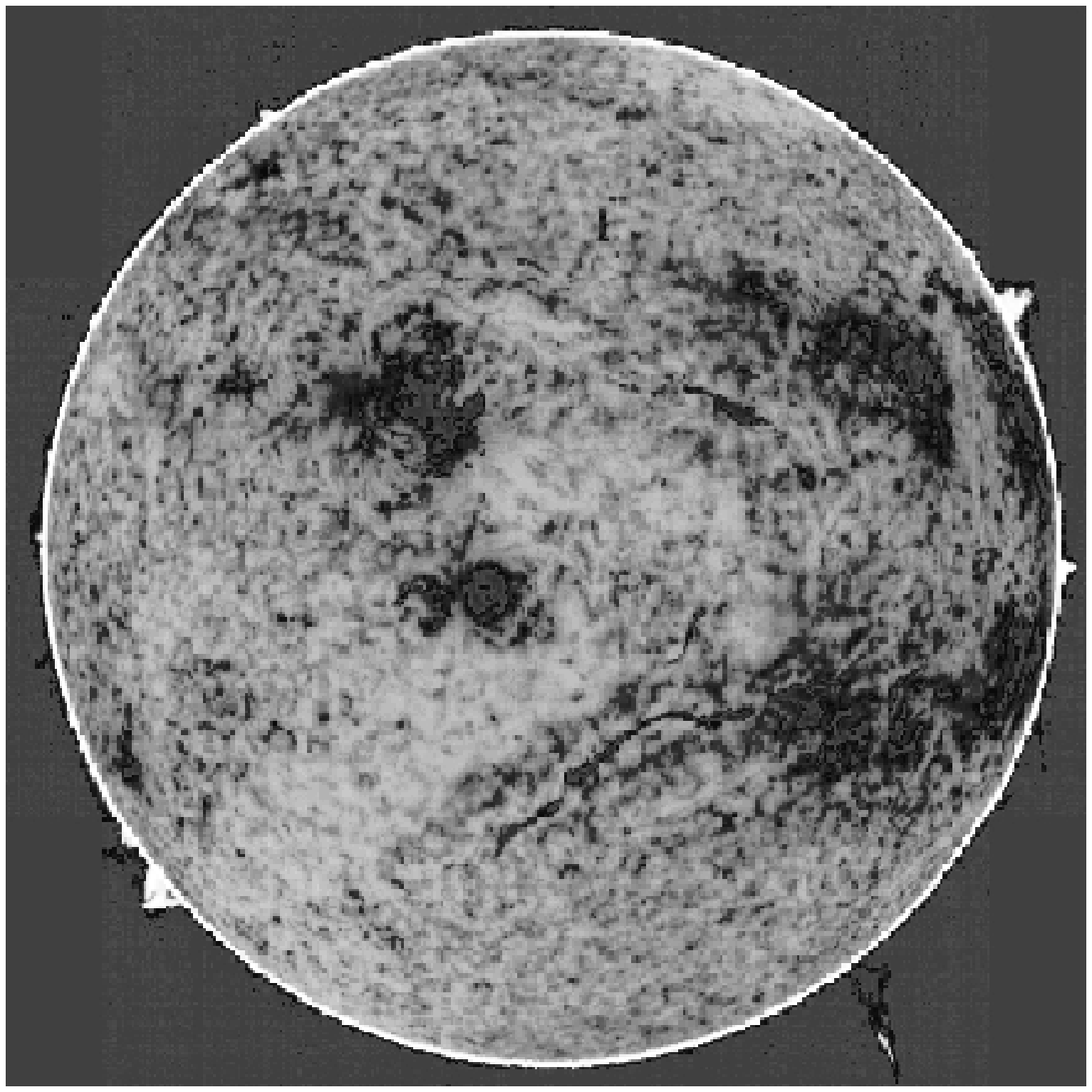}{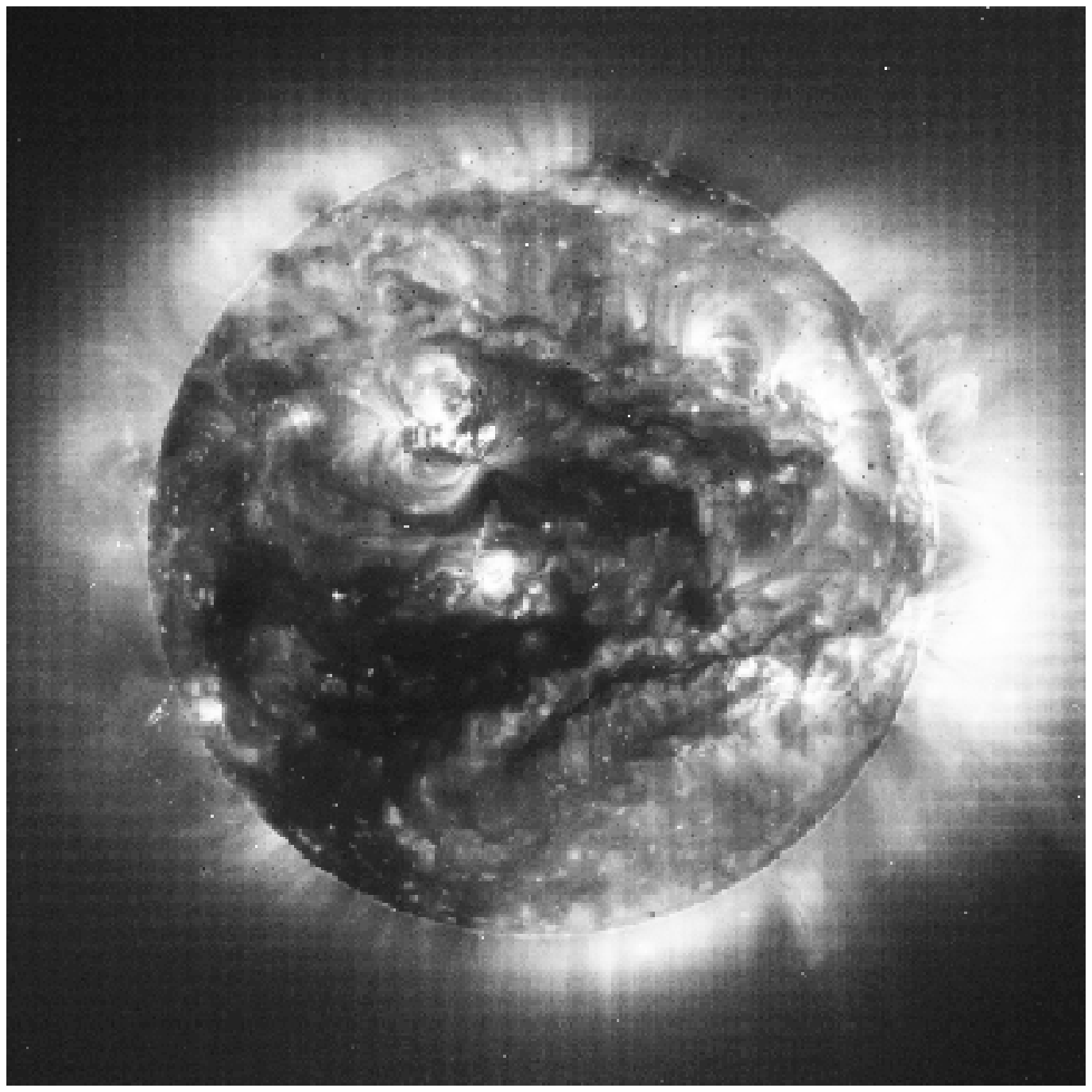}
\plottwo{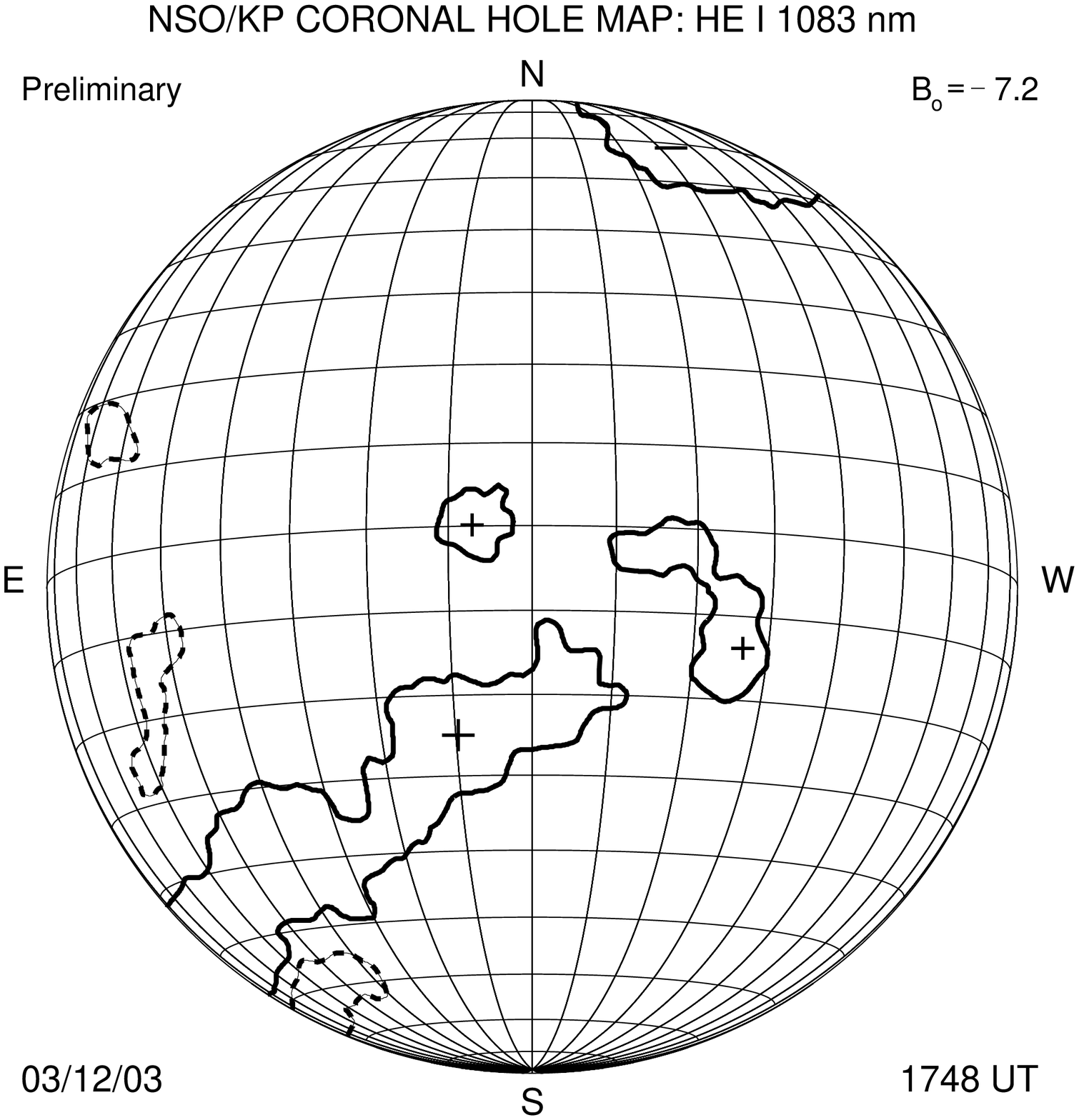}{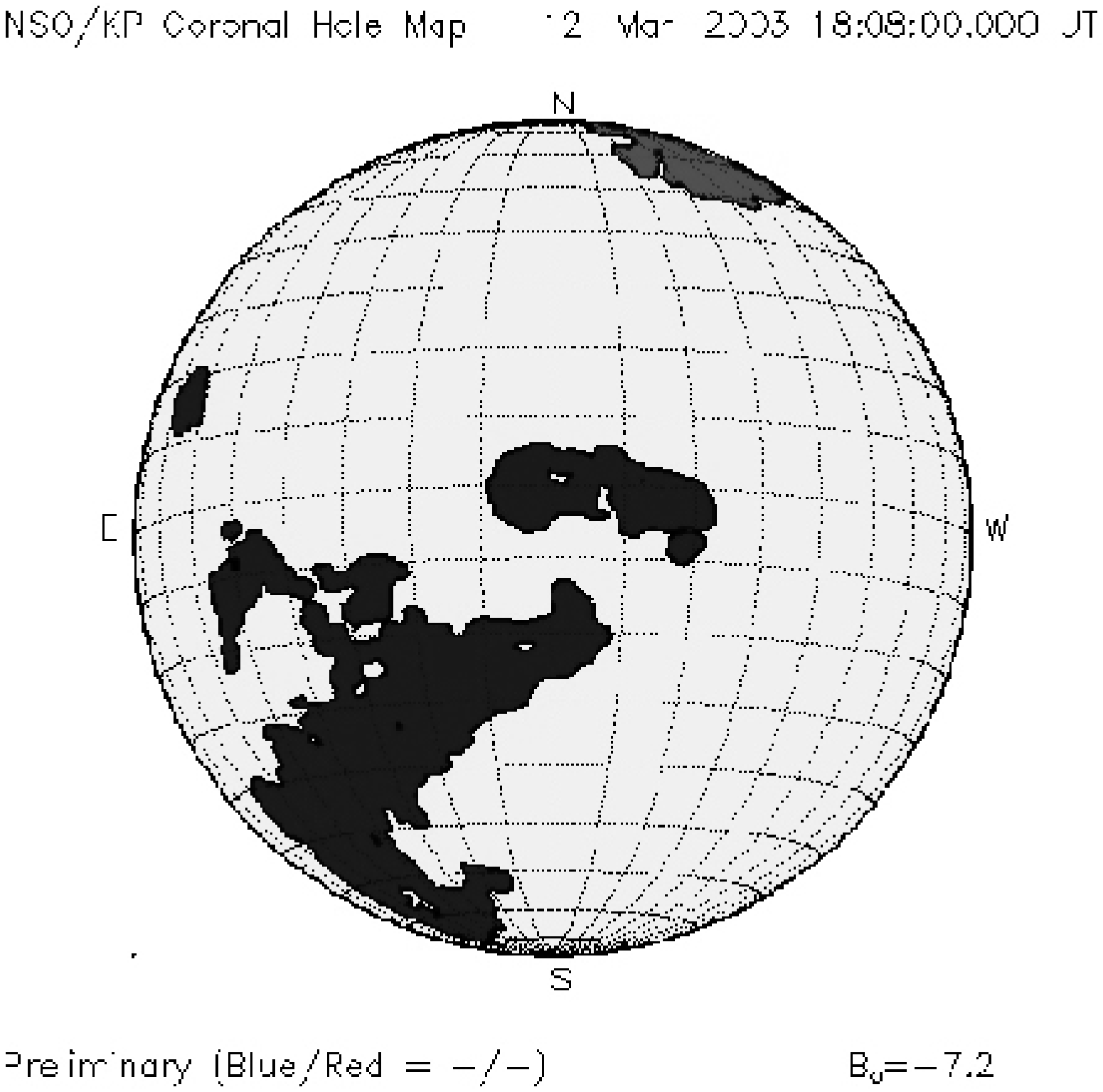}
\caption{An example when the HD (bottom, left) and AD (bottom, right)
coronal hole images disagree. KPVT He~{\footnotesize I} 1083~nm
spectroheliogram (top, left), and EIT 19.5 nm Fe XII emission line image
(top, right) observed on March 12, 2003 are also shown.}
\end{figure}

Starting with the average helium image, the first step 
is to retain only values above $1/10$ the median of positive values
(based on the mean values shown in Figure~2). 
After removing the negative values, the candidate coronal hole regions 
become clearly visible (see Figure~3c). The next two steps are to spatially 
smooth the remaining positive values and set all positive value pixels 
to a fixed value. The resultant image is then smoothed with the 
morphological image analysis function Close \citep[e.g.][]{mich2001}
using a square kernel function as the shape-operator.
Shown in Figure~3d, the Close function fills the gaps and connects associated 
or nearby regions to define the areas of the coronal hole candidates.
The regions that are too small to be coronal holes are then removed
\citep[see][]{KHarvey2002}. The spatial image pixel positions of 
non-zero spatial points of the resultant image (e.g. Figure~3d) are used
to fill the corresponding pixels in the image that was created after 
the first step (e.g. Figure~3c) with large values (e.g. $10^4$). 
After being spatially smoothed, the large 
values result in the filling of small gaps and holes. 
The resultant image is then smoothed with the morphological image analysis 
function Open \citep[e.g.][]{mich2001}, where the Open function removes small
features while preserving the size and shape of larger regions (see Figure~3e). 
For more discussion on morphological image analysis and standard algorithms 
for segmenting images see \citet{jah2004}.

\begin{figure}[!t]
\plotone{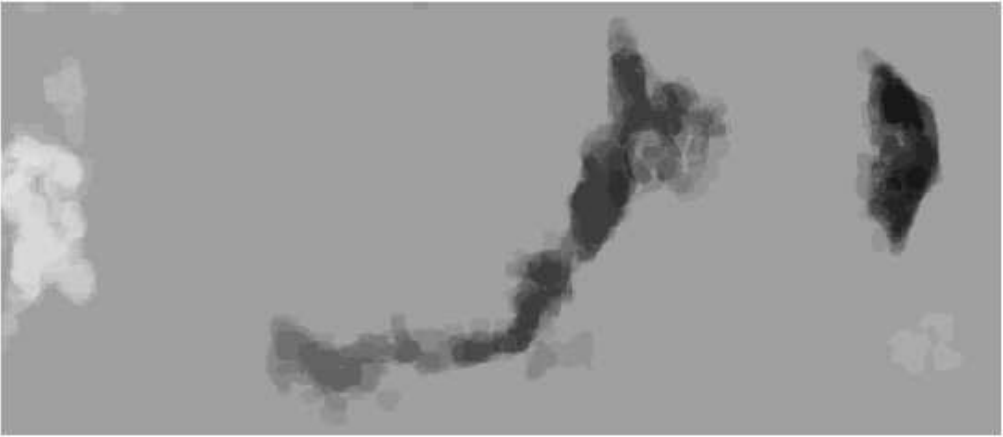}
\caption{Subregion of an automated coronal hole synoptic map for Carrington 
rotation 2005. The gray-scale reflects the polarity (positive and negative 
are represented by white and black respectively) and the number of coronal 
hole detections at a given latitude and longitude. The central region is the 
same coronal hole exhibited in Figure~1 and~3.}
\end{figure}

The primary goal of the steps so far has been to consolidate and define
the boundaries of candidate coronal hole regions. The next step is to use
the average magnetic image data to determine
the percentage of unipolarity of each candidate region.  
Candidate regions that have a percentage of unipolarity lower than the
thresholds (delineated by the dashed lines in Figure~2) are removed. The final 
step is to number the coronal hole candidates and apply the sign of the 
region polarity (e.g. Figure~3f). Candidate regions with mean unipolarity values that 
fall within the dashed and dotted lines depicted in Figure~2 are considered provisional. 
These regions are distinguished in the final output image by adding a 
fractional value (e.g. 0.5) to the region number value of the image pixels for the 
given coronal hole area.

\subsection{Coronal Hole Map Comparison}
In general the coronal hole candidate regions detected by the AD algorithm
agree well with the HD images. For the 11 years compared, the automated coronal 
hole areas have approximately 3\% less area than the hand-drawn 
regions between -50 through 90 degrees latitude. The difference increases
for the latitude band between -90 and -50 degrees. However, the increase is 
believed to be the result of incomplete observations 
used in the average images (i.e. the KPVT full-disk observations always 
began in the north). Comparisons between the HD and AD coronal hole images
are shown in Figures~1 and~4. The two images agree for the July 14, 2003
example shown in Figure~1. However, comparisons with only the HD maps have the 
limitation of not comparing directly with coronal measurements. Figure~4 highlights this 
difficulty. For this example, the AD map is in better agreement with the EIT image. 
A detailed comparison with other coronal hole detection 
methods \citep[e.g.][]{detom05,Malan05} is planned. These comparisons will undoubtedly 
result in further refinement to the presented AD algorithm.

\subsection{Coronal Hole Synoptic Maps}
In addition to creating coronal hole images, these individual images are combined 
to create synoptic maps for fixed Carrington rotation or daily synoptic maps.
Each pixel value of the synoptic map represents the number of detections that a 
coronal hole is observed for that location.
A subregion from an example Carrington map is shown in Figure~5.
These maps reflect the estimated coronal hole boundary variations as a result
of temporal evolution of the region in addition to the quality of the detection.
The new synoptic maps will be publicly available via the NSO digital library 
along with the daily AD coronal hole images.

\section{Acknowledgments}
The coronal hole data used here was compiled by K. Harvey and F. Recely 
using National Solar Observatory (NSO) KPVT observations under a grant 
from the National Science Foundation (NSF). NSO Kitt Peak data used here are 
produced cooperatively  by NSF/AURA, NASA/GSFC, and NOAA/SEL. The EIT images 
are courtesy of the SOHO/EIT consortium. This research was supported in part 
by the Office of Naval Research Grant N00014-91-J-1040. The NSO is operated 
by the Association of Universities for Research in Astronomy, Inc. under 
cooperative agreement with the NSF.

\end{document}